# The Study and Optimization Of Production/Fermentation Processes In Biofuel Production

Amardeep Singh

**ABSTRACT**

The production process involved in the creation of biofuels consists of a number of operations and steps that require a meticulous understanding of the parameters and metrics. The production techniques again differ depending on the pre-treatment systems, source material, the methods used for extraction, types of nutrients used, cell cultures employed, time undertaken and temperature.

Due to the strategic and crucial role that bioethanol holds in supporting the energy demands of the future, it becomes important to run such processes to a highly optimized extent. One of the frontiers of leading such optimized designs is by studying the data from the production processes, formulating design experiments from said data and correlating the results with the parameters using analytical tools.

Data analytical tools have found use in nearly all spheres of the industry- from determining compatibility of land for growth characteristics to optimizing supply chain systems for fuel transportation. The scope of the project would thus be to first uncover data about the production procedures from literature and develop a predictive model of the data which can provide greater insights in the determinants and their effects on the results. While the case examples analyzed relate to bioethanol mostly, an additional analysis has been performed for data on biodiesel.

Coupled with confirmatory methods such as Principal Component Analysis, researchers can help narrow down the extent or degree to which the parameters affect the final outcome and even configure inputs that may not play a definitive role in greater outputs.

The project first tackles through some conventional case studies involving biofuel production using an FIS(Fuzzy Interface System) and provides certain insights into the ways in which fuel yields can be enhanced depending on the particular cases. For the purpose of analysis, tools such as MATLAB, Python and WEKA have been employed. Python and WEKA have been used extensively in building principal component analysis reviews for the purpose of this project while MATLAB has been used for building the FIS models.

**TABLE OF CONTENTS**

# 1. AN INTRODUCTION TO PARAMETRIC PRODUCTION TECHNIQUES

Algorithmic design of experimentations have helped pave the way for industries and manufacturing sectors to restructure and rethink the way they expend resources to produce desired products at massive scales. Numerical analysis techniques such as the Taguchi method, Signal to Noise Ratio test and ANOVA have allowed industries to further narrow down their approaches in shaping production system around appropriate conditions while also cutting down on costs and resources. *(Ezzatzadegan, Morad, & Yusof, 2016)*

The use of such statistical techniques can also be observed extensively in the process control and quality control sectors where input data is employed in reliability studies, sampling cases and building safety models for dynamic systems. Parametric design methodologies take into account the central fact that the parameters alone influence the output in the model to be built and can increase as the datapoints grow more complex.

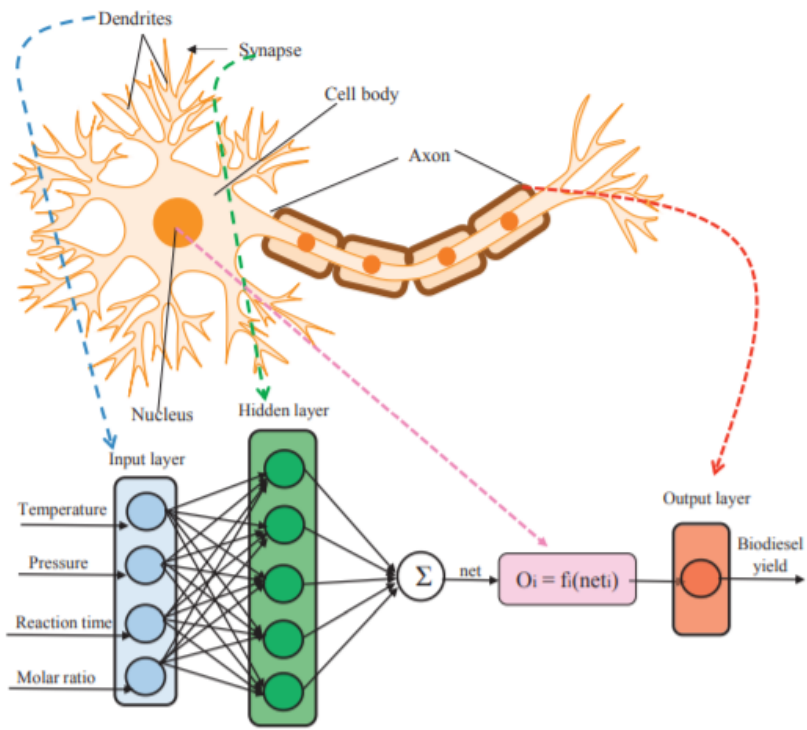

*Figure 1.1-Schema of a Neural Network (Farobie, Hasanah, & Matsumura, 2015)*

Artificial Neural Networks(ANN) are one such example of a semi parametric design case where the input data is cleaned, filtered, prepared and fit against the output using a prescribed number of smaller algorithms. Mimicking the structure and function of a neural schema, ANNs can have multiple hidden layers with initial biases and even smaller algorithms such as gradient descent, Levenberg-Marquardt, scaled conjugate gradient and resilient backpropagation. *(Nielsen, Larsson, van Maris, & Pronk, 2013)*

Before being processed by such algorithms, the data is often standardized or normalized to better help the system learn and train the models. This is done in order to eliminate system complications that can arise when working with datatypes that differ in units and ranges between maximum and minimum values.

While there may not be a singular function that can fit all sorts of data, the essence behind parametric studies is to better understand the inputs and leverage the influence that each one has on the output. Accuracies of models are highly influenced by the number of iterations,

epochs(periods of runs), training to testing dataset ratio, number of layers and the general input to output ratio.

Industrial applications of ANNs have been a technical leap for manufacturing centres which is a great shift from the generic PID(Proportion-Integration-Derivative) control systems for handling highly convoluted procedures such as cement manufacturing and drug design. Such processes are highly benefitted by the ability of algorithms to predict and perform necessary changes that have to be implemented as a result of external variabilities and altering production cycles. Adding filters such as the Kalman filter (Sohpal & Singh, 2015) can help capture yields for multiple complex outputs dependent on various inputs.

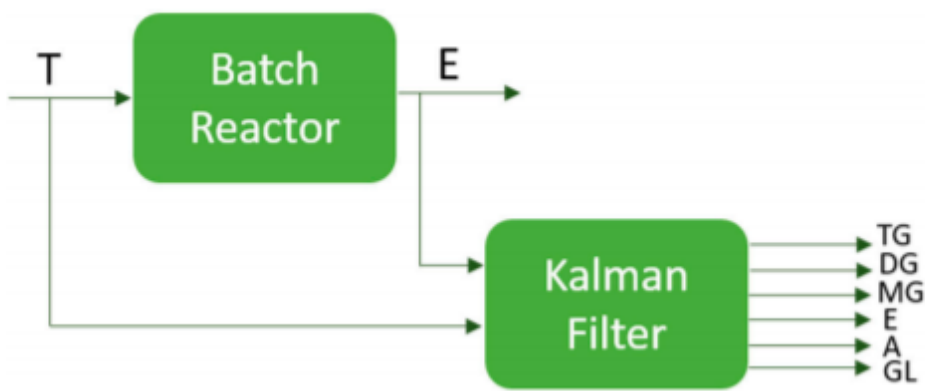

*Figure 2.2-Algorithmic Design Filter (Sohpal & Singh, 2015)*

In the realm of biofuel production, such process design methodologies have been used to better interpret the conditions that give rise to better yields as well as comparing the economic advantages of certain unit operations along with various feedstock materials. ANNs are particularly invested for process data where deep learning strategies are required, such as for design cases with voluminous rows and columns. *(Cardona & Sánchez, 2007)*

Due to the self learning nature neural network methods and fuzzy logic systems, companies can generate adaptable models that can alter as a result of the production parameters and are not confined to a singular definition of algorithm generation.

The choice of which algorithm would yield the best accuracy is however left to the user and can influence the resultant weights obtained for all factors studied in a typical production process.

## 2. THE FUZZY LOGIC ADVANTAGE

Created by mathematician Lotfi Zadeh, fuzzy mathematics transcends concepts in truth categorizations, analytical sciences and degrees of variability. Considered to be a mathematical branch that deals with multiple states of outputs, fuzzy mathematics gave rise to the use of logical truth systems to connect data inputs and outputs.

The objective with fuzzy mathematical systems is directed towards representing crisp data in terms of 'fuzzy' or range-based values. Once these fuzzy data points have been connected with the output, which too has been converted to fuzzy ranges, the system 'defuzzifies' the functions that correlate to the output and produces a crisp numeric value.

Unlike more conventional parametric models, fuzzy logic involves breaking down the inputs into a series of categorical variables where they can lie between a range of values. These variables are defined to illustrate the effects they produce on the output which is again represented and reformatted as a sequence of ranges.

On the conversion of the numeric data to categorical variables, rules are defined and inserted into the system that connect the input to the output, albeit now, as linguistic ranges. These ranges are referred to as membership functions which denote that the linguistic significance of the function lies between two specific numbers. The membership functions can be reflected using a number of formulations including triangular, gaussian, trapezoidal and more specific, singular designs. On the creation of a considerable number of rules, the system can thus generate a fuzzy understanding of how outputs are influenced by the inputs.

To make the models more accurate, more membership ranges can be implemented as well as bringing in more definite functions that better capture the values. When compared to more conventional process dynamic methods that preprocess the data on the same scale and range, fuzzy logic designs benefit by not having the need for such massive formulations.

Due to their potential in tracking out variations and dealing with process outliers, fuzzy logic based controllers are often implemented alongside standard units in industries, often after learning and interpreting the data immensely. These can be considered to be predictive algorithms that track changes in data by comparing testing to training ratios, rather than using integrative and derivative functions to study how parameters will change. *(Singh, et al., 2013)*

To further augment the accuracies of the controllers, the fuzzy interface systems are linked to neural network schemas to create a neuro-fuzzy controller. It is important to emphasize that the use of such arrangements in industries have helped better predict and scale output variations, especially towards the mean. Fuzzy logic derives its use from multi level criteria decision making based on a system of simple logical conditions connected to the output.

The degree with which the elements in the logical conditions confer to the output are decided by the user and can be adjusted to great produce great accuracies when studying similar input dimensions in future studies.

## 3. MODELLING IN BIOFUEL PRODUCTION

Biofuel production has become a subject of significant interest in answering the question to rising energy needs as well as combating widespread emissions which are detrimental both to the environment and to human health. It is thus highly imperative for industries to structure production methods that promise great yields with reduced variations from the standard operating conditions.

ANNs equipped with tools such as Response Surface Methodology(RSM) have been primarily used to design limiting systems in manufacturing industries that keep parameters such as pH, temperature, inoculum concentrations and reaction times to their suitable ranges.

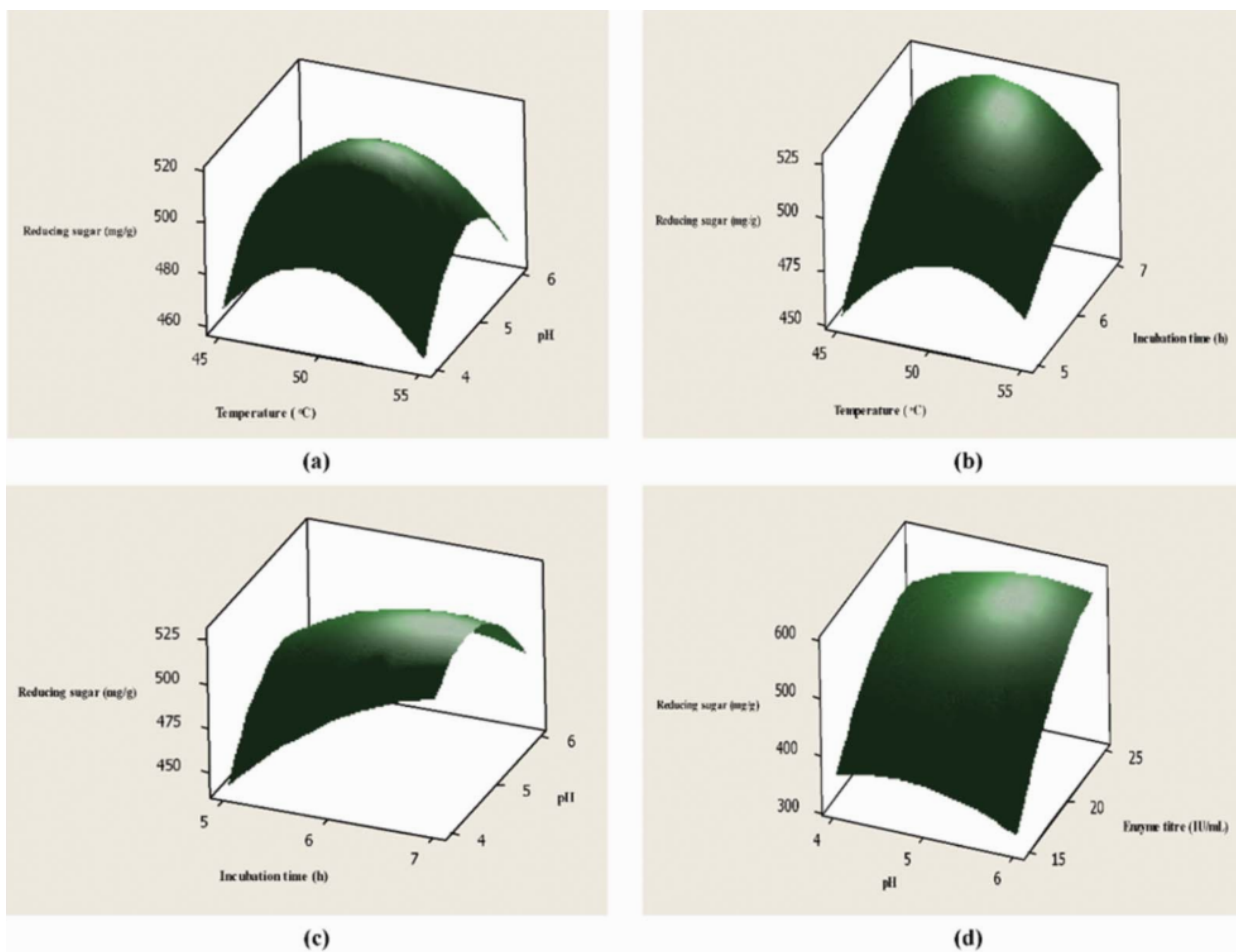

On-site sensors further provide industries with the data necessary to correlate production parameters with process yields, helping save time and resources. Deep learning models too have been studied for predicting the yield of different types of feedstocks, leveraged against economic conditions and constraints. *(Nielsen, Larsson, van Maris, & Pronk, 2013)*

Such systems have also found use in developing modular techniques for configuring enzyme concentrations in biofuel extraction from lignocellulosic materials, particularly for breakdown and fermentation. However, a vast majority of training softwares offer different advantages to the modelling step, which is defined by the tools and packages available.

Conventional methods of predicting process parameters for biofuel production take example from older industries that employed tuned PID systems, often with multiple units running simultaneously. These variables are first represented as transfer functions in the software system and then inputted to the PID framework. The system generates plots of the parameters, allowing users to track deviations and strategically alter the transfer function coefficients to bring the system to stability. *(López-Zapata, et al., 2017)*

Stability becomes an important fact to consider for biofuel production due to the extraneous possibilities of the system to output smaller yields due to unwanted, uncontrolled or unexpected changes caused both internally and internally. In real time monitoring systems, these corrective methods, such as the Kohen Coon fine-tuning method, have to be inputted almost instantly before new values are captured from the production floor.

Such models also take into account delays that can develop between the values being measured in the production unit and the value reaching the control unit by introducing zero order holds. The equations for these 'holds' are created using z transforms of transfer functions which require complex mathematical simplification to single order equations. Commonly used algorithms that use such equations include the Deadbeat algorithm and the Dahlin algorithm in industries.

The complexity is further compounded to a great degree in multivariate design systems where parameters can have a confluence effect on the output and some may not even influence it to any degree. As the order of the parameters increases, adjusting the new coefficients(of orders often above 10) becomes extremely arduous.

Biofuel production, particularly, employs batch reactors in a consecutive fashion which allows control teams to incorporate previously obtained values into system models before a new batch operation is initiated. However, during the production phase, such updates have to be fed instantaneously in the model so that they can generate possible future values. A common issue that Fuzzy Logic systems help tackle are cases of redundant inputs generating multiple outputs. As the user generates the rules to capture these redundancies and sets boundary conditions for each instance, predicting them becomes simpler. *(López-Zapata, et al., 2017)*

4. **THE CYCLE OF DATA MODELLING**

The conventional model of data modelling for predictive and preventive studies first involves a well defined data collection procedure. Sensors and actuators are placed at key points throughout the entire production floor which capture the values for measurable parameters.

New data is derived from the existing data by using feature engineering techniques. The data is then cleaned and preprocessed to understand the distribution of values often using log scales and histograms. (Sohpal & Singh, 2015)

Once the data has been cleaned, it undergoes a preprocessing step to eliminate the differences in units between the variables using normalization or standardization. To further interpret the completer interaction between the variables to the output, a graphical PCA(Principal Component Analysis) is created.

The purpose to the PCA is to reduce the effects of a large number of columns to a simple dimension of coefficients(as determined by the user). The PCA scores obtained from these become the points to be plotted on the graph.

Depending on the statistical nature of the columns(numerical or categorical), they may undergo another step to be equalized. If the output to be modelled is a numerical value, regression techniques are employed, otherwise, classification becomes the preferred tool. In the context of fuzzy logic, the steps in model building are less intensive on preprocessing the values and more concerned with defining truth functions for the input and output. *(Nielsen, Larsson, van Maris, & Pronk, 2013)*

## 5. CHALLENGES IN MODERN MODELLING

Apart from the general uncertainties that can arise in the datasets from unaccounted covariates or outliers, the rate at which coefficients are updated as per the models can have drastic effects on accuracies. In a standard industrial setting that employ PLC SCADA models built from theoretical approaches, the possibilities of redundant data becoming a part of the setup is not uncommon. Other errors can arise from incorrect or erroneous machines and sensors which may not be calibrated properly to monitor system settings properly. Another issue in modern modelling setups can arise from failing to procedurally connect the actual inputs that influence the output. Covariance studies alone cannot determine the degree to which both may be linked.

In the case example of the yield being optimized as per the methanol to oil ratio, (Farobie, Hasanah, & Matsumura, 2015) it seems astounding as to how it has a major effect on the output when removed. However, if it is removed along with the reaction time or temperature, the model becomes less accurate.

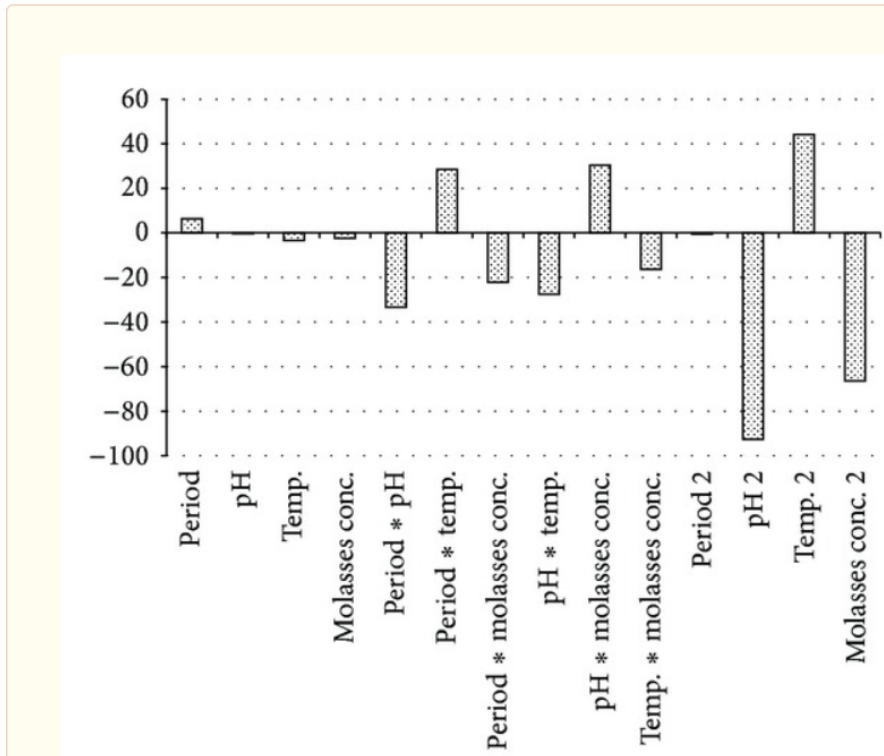

*Figure 5.1-Influence of various parameters and combined effects on output( El-Gendy, Madian, & Amr, 2013)*

Another complication happens to be the number of membership functions that are used to capture the values in ranges. Excessive number of membership functions may make the model more precise but can lead to overfitting, rendering it inefficient for new datasets in the future. If the membership functions are reduced, the model can generate redundant output values for different inputs. As a result, it becomes important to generalize special cases and rules to the model. *(Singh, et al., 2013)*

Artificial Neural Networks when combined with fuzzy logic systems can help meet accuracy requirements, provided there is a good training to testing ratio provided for the data. In the case examples that are discussed in this paper, the experimentations were able to arrive at a conclusion about the best algorithms to use after careful consideration of the parameters using a central component design matrix or by studying RSM plots.

# 6. INSIGHTS FROM THE INDUSTRY-PRELIMARY ANALYSIS AND EXPERIMENTATIONS

A simplistic principal component analysis of the inputs and outputs can help reveal how they correlate with each other. The sample literature from which the data has been obtained have been used to illustrate this here. To perform the principal component analysis, the tabular data is first normalized to yield working values which then become suitable for modelling.

The following graphs show the plotted points for the ethanol yield factored by the inputs for various cases, carrying from the objective of the individual paper. The directions in which the arrows point indicate the proportional or inverse relationship between the corresponding variables.

It has been noted that design experimentations are devised using softwares like Qualitek which narrows down the possible number of combinations for optimizing the production outputs. This play an important role in removing recurring data steps and avoid repeating calculations for the learned model. Eventually as the model learns the experimentations better, an optimized design standard can be obtained. (Nielsen, Larsson, van Maris, & Pronk, 2013)

Research has suggested that sugar derived from molasses is a cost effective feedstock source, as a result of its higher sucrose content. Optimization studies have favoured S.cerevisiae as the main industrial microorganism for creating ethanol as a result of its high sucrose efficiency.

In the literature discussed by Rivera, et al.,(2015), sugarcane baggase was processed using a pretreatment step involving sulphuric acid with various solid loadings and acid concentrations. The effects of all these parameters was noted along with reaction time to determine the digestability and the glucose yield. This is a standard MIMO(multiple input, multiple output) neural network where it was observed that the acid concentrations were producing insoluble hemicellulose which decreased with an increase in sulphuric acid. Higher solid loadings had a reasonable imapct on the digestablity of the biomass as well as hetereogenous reactions catalyzed by the cellulase.

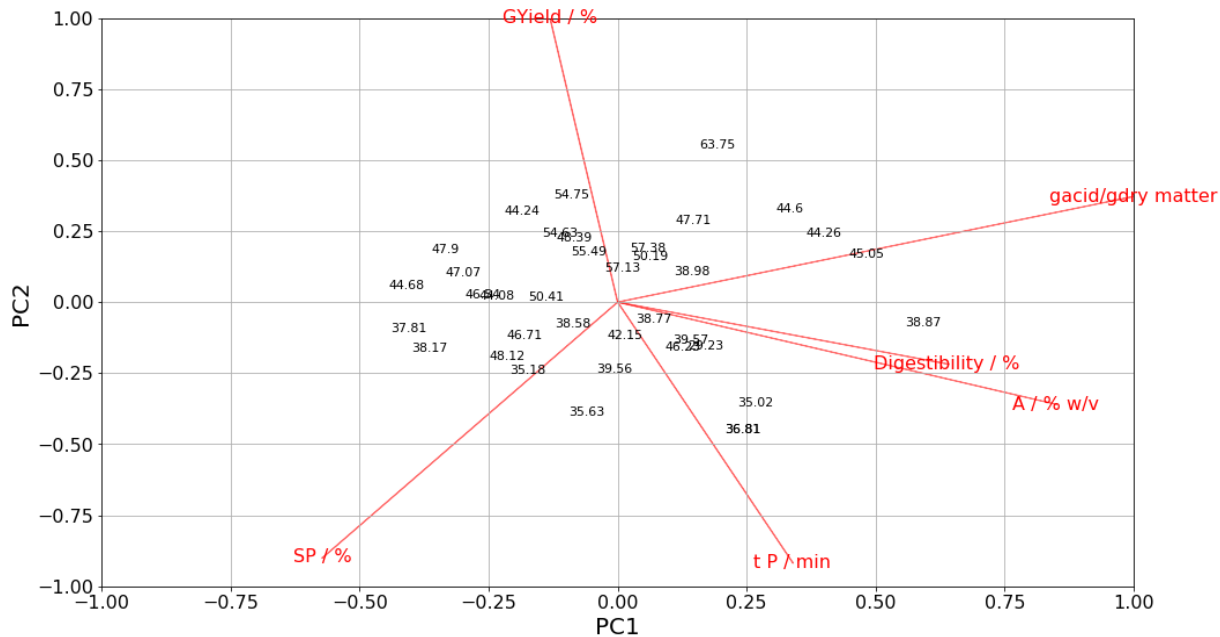

*Figure 6.1- PCA Simulation Results from data obtained from (Rivera, et al., 2015)*

In the above graph for example, the normalized data points for the output yields have been plotted as numbers to show how the output yields are related. The inputs t P/min(time) and GYield % are pointed in opposite directions, indicating that they may in fact share an inverse relationship for this particular production experimentation. The axes represent the 'scores' of the principal component analysis which reduces the modular equations for the normalized inputs into a set of linear coefficients.

In the next case example studied in Adnan, Suhaimi, Abd-Aziz, Hassan, & Phang, (2014), experimental factorial design and response surface analysis were used to optimize ethanol production. The parameters A,B,C,D in the PCA chart below indicate the initial pH, substrate concentration, organic nitrogen concentration and salt content respectively.

The results indicate that glycerol fermentation for ethanol production is best suited using E. coli SS1, especially when considering a batch reactor. It is imperative that more than 90% of the initial glycerol be converted into ethanol which can be achieved by building predictive models. This particular case uses anaerobic conditions. Parametric studies help reduce the number of feasible inputs to just 4 which simplifies the model further. Experimentally it has been observed that as the glycerol concentration increased from 20 g/L to 45 g/L, ethanol yields doubled.

A fermentation period of 72 hours allows for cells to proliferate to109 CFU/ml. Higher cell densities help contribute to better yields even when glycerol concentrations are high. The direction of the arrows B and D indicate that they may have an inverse relationship while A and C are closely related.

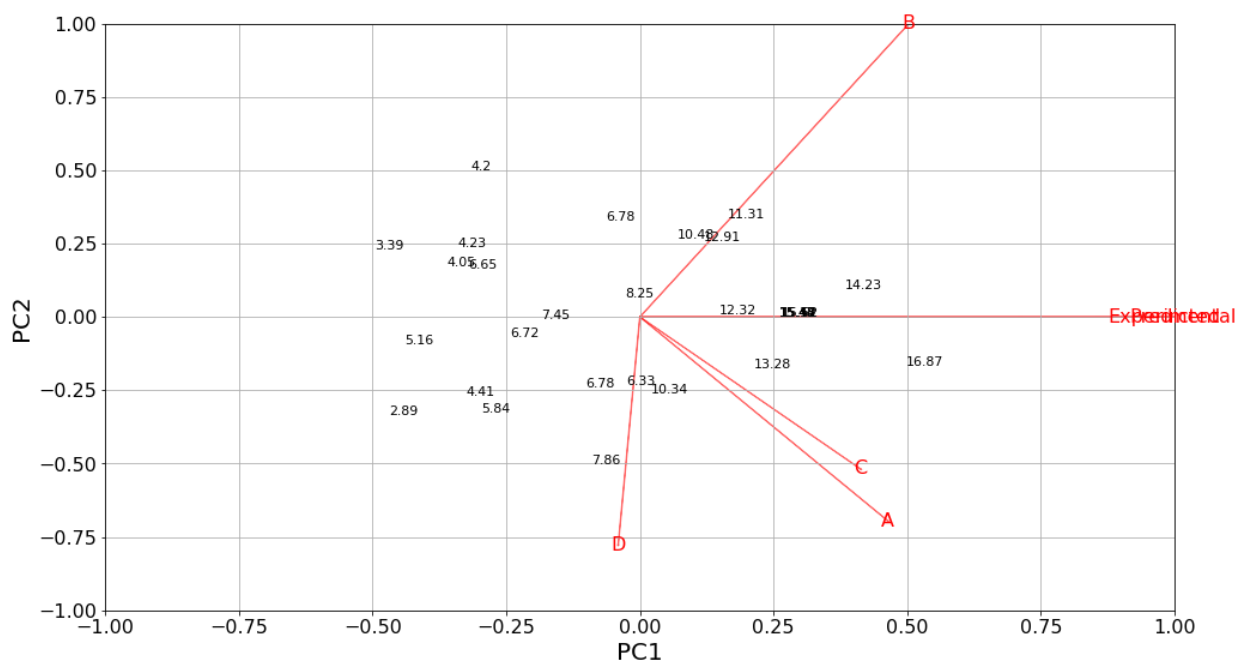

*Figure 6.2-PCA Simulation Results from E.coli Strain Yield(Data obtained from Adnan, Suhaimi, Abd-Aziz, Hassan, & Phang, 2014)*

In the next example, we take a contrasting look at the parameters for maximizing yields from lignocellulosic biomass. The complex nature of the steps involved in the conversion of cellulose to its monomeric sugars, necessitates the requirement for accurate predictive models.

This experimentation as conducted in Sherpa, Ghangrekar, & Banerjee(2017) discuss the composiste design for biomass that undergoes pre-treatment, saccharification and fermentation. The modifications in terms of functional groups were studied using Fourier transform infrared (FTIR) spectroscopy. Researchers used the KBr pellet technique where the samples were dried and pressed into a potassium bromide (KBr) disc.

Fermentation was conducted using S. cerevisiae and as per the PCA chart, we can see that while Temperature may not have a major effect on yields, solid loading and pH have a partially inverse relationship to the output(Reducing Sugars). Data does suggest that the yields may be higher than acid pretreated and hydrolysed sugarcane tops.

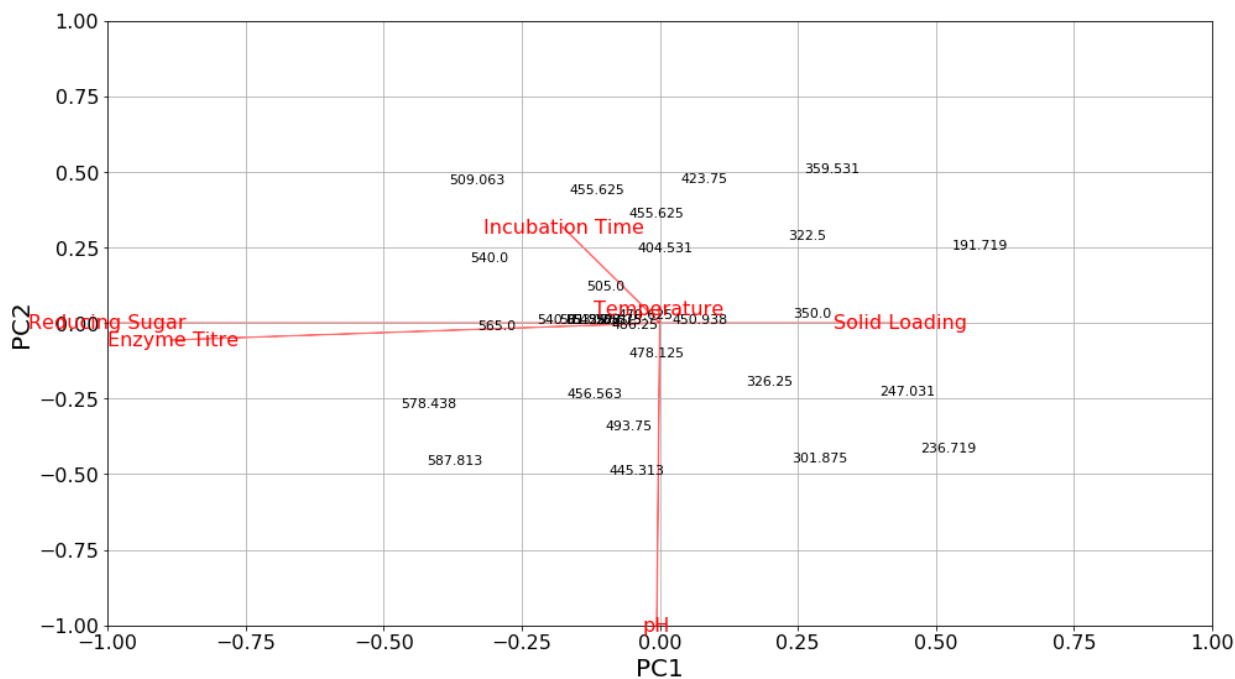

*Figure 6.3- PCA Simulation Results from experimentations (Data obtained from Sherpa, Ghangrekar, & Banerjee, 2017)*

The next case example discussed in (El-Gendy, Madian, & Amr, 2013) used an RSM and CCFD matrix for modelling four variables- incubation period, initial pH, incubation temperature, and molasses concentration. The experiment again used locally isolated Saccharomyces cerevisiae Y-39 with maximum yields predicted to be 253 and 255 g/L, respectively.

The design also helped narrow down optimal conditions for experimentations to be centred at 71 h, pH 5.6, 38°C, molasses concentration 18% wt.% for 100 rpm. The experimentations used additives such as ammonium sulphate, magnesium sulphate and monopotassium phosphate. pH and temperature were adjusted to allow for maximal growth of the cultures. Cultures for the medium were created in the medium for 48 h at 30 degree C in a shaking incubator at 150 rpm. Cells were then resuspended in another solution following cleaning with saline water.

Liquid chromatography was used to determine the type of molasses present. Experimental runs were carried out to a full 24 hour factorial design with the four variables. The number of runs was interestingly determined by the number of independent variables. Under RSM methodologies, the variables were designated levels(ranging from -1 to 1). The PCA plot for the variables obtained from Python show that initial pH and incubation temperature have an inverse influence on the yield. Similar relationships can be seen for the molasses concentration and the incubation period.

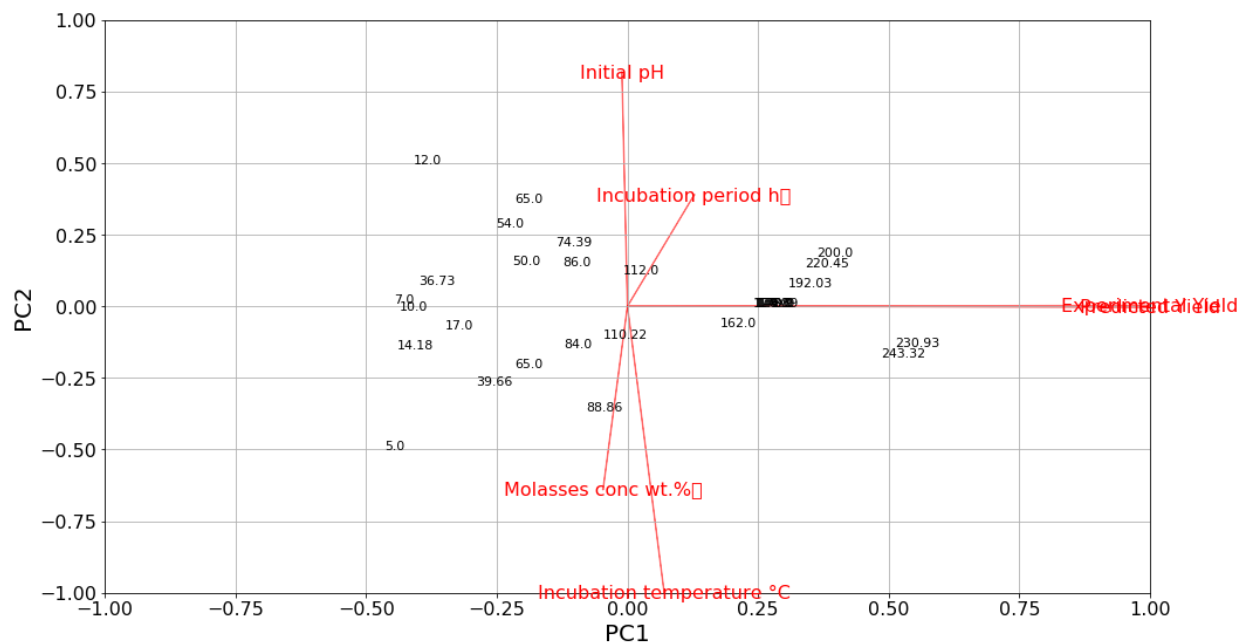

*Figure 6.4- PCA Simulation Results from experimentations(Data obtained from El-Gendy, Madian, & Amr, 2013)*

To further understand the use of variable designs in industries, we take a look at the data obtained from (Farobie, Hasanah, & Matsumura, 2015) which shows the variable effects on non-catalytic biodiesel production from supercritical methanol (SCM) and supercritical ethanol (SCE).

A spiral reactor was used for this experimentaiton which was formatted using an artificial neural network (ANN) model in order to predict biodiesel yield. With a low error rate and mean square error, the model can become vulnerable to overfitting and fail to properly predict future input values. The model showed that the highest yield for biodiesel was determined at an SCM concentration of 1.01 mol/mol corresponding to the actual biodiesel yield of 1.00 mol/mol.

This was obtained at 350 °C, 20 MPa within 10 min. Similarly the highest possible yield for biodiesel from SCE was 0.97 mol/mol achieved at 400 °C, 20 MPa within 25 min. The PCA plots obtained show how using a spiral reactor can have varying effects on the correlational statistics between the inputs. In the plots obtained from Python, all variables for SCM and SCE are directed in the same direction as the yield showing a net positive relationship.

The change in direction from left to right in Figure 6.5 to Figure 6.6 are indicative of the relationships the PCA scores have with the standardized inputs and outputs. This still shows that having a higher pressure, chemical to oil ratio and reaction time promise higher yields.

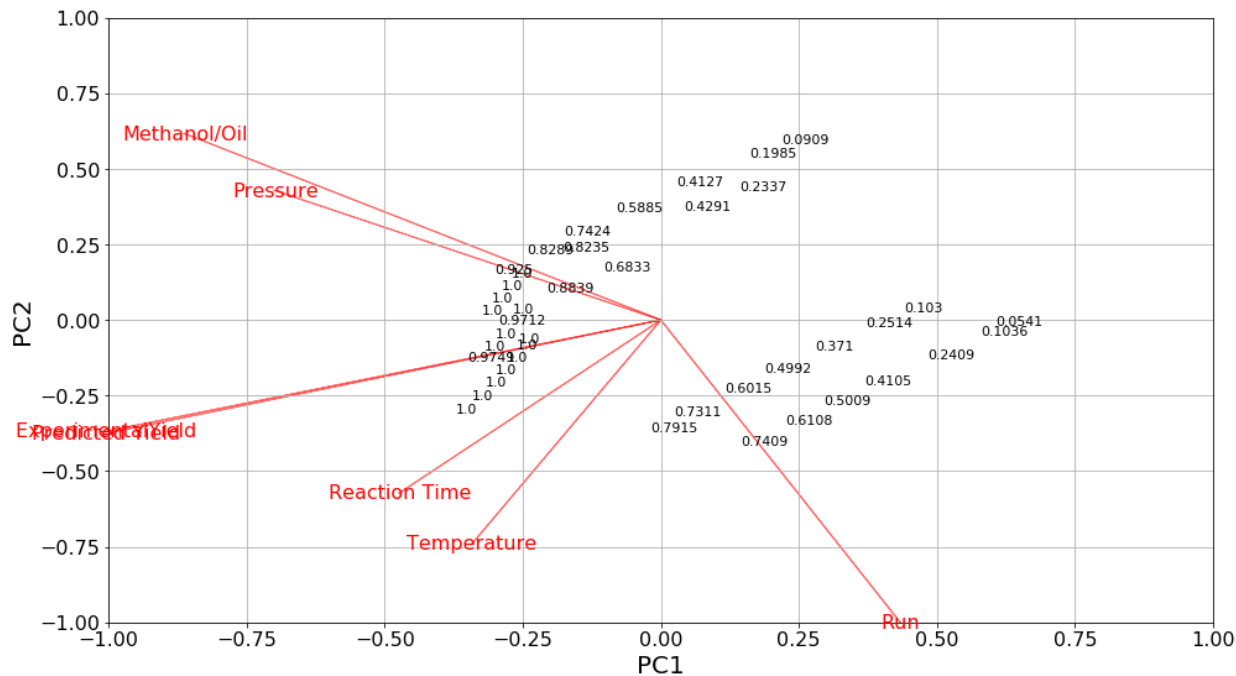

*Figure 6.5- PCA Simulation Results from experimentations(Data obtained from Farobie, Hasanah, & Matsumura, 2015)*

The experimentations show that higher yields can be achieved with better optimal processing conditions that are the cornerstone of model predictive methodologies. When dealing with variables that are more intrinsic such as reaction time or pH values combined with yeast inoculum concentrations and nutrient concentrations, such models help deal with minor changes to a better degree than standard predictive steps. This has been discussed extensively in the model cases in the following sections.

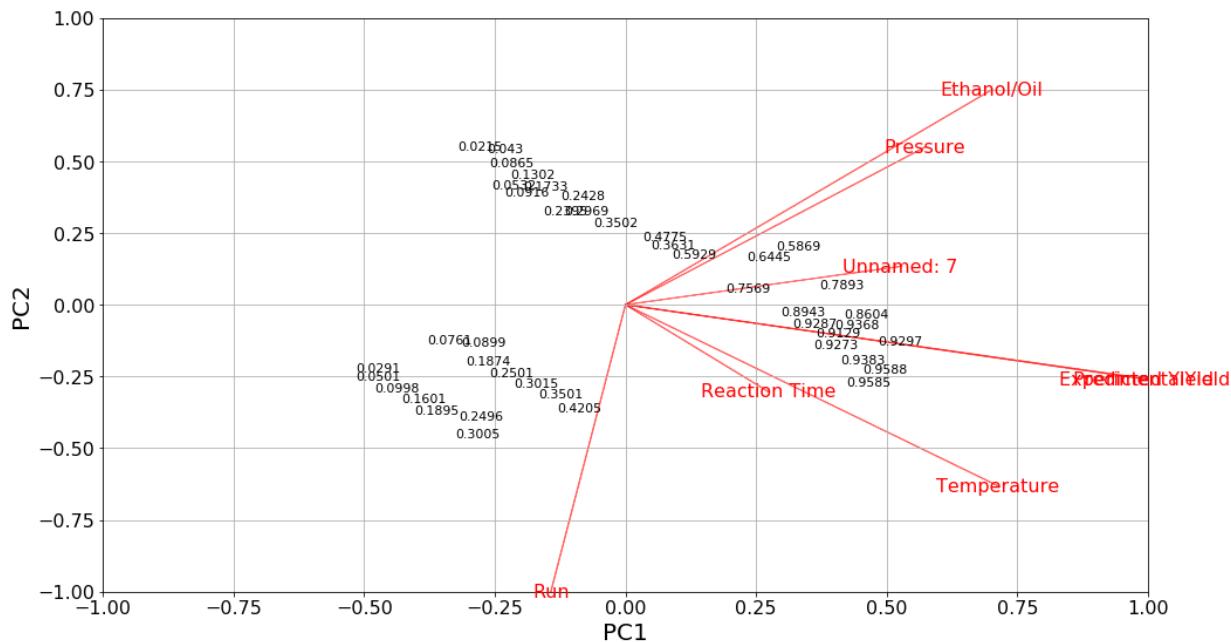

*Figure 6.6- PCA Simulation Results from experimentations(Data obtained from Farobie, Hasanah, & Matsumura, 2015)*

## 7. MODELLING DATA-FUZZY MODELS

Modelling in the fuzzy interface system first begins by specifying the number of inputs and outputs in the FIS window. The ranges of values are inputted as well as the type of membership functions they will have. This could range from triangular, trapezoidal, pentagonal, gaussian and so on.

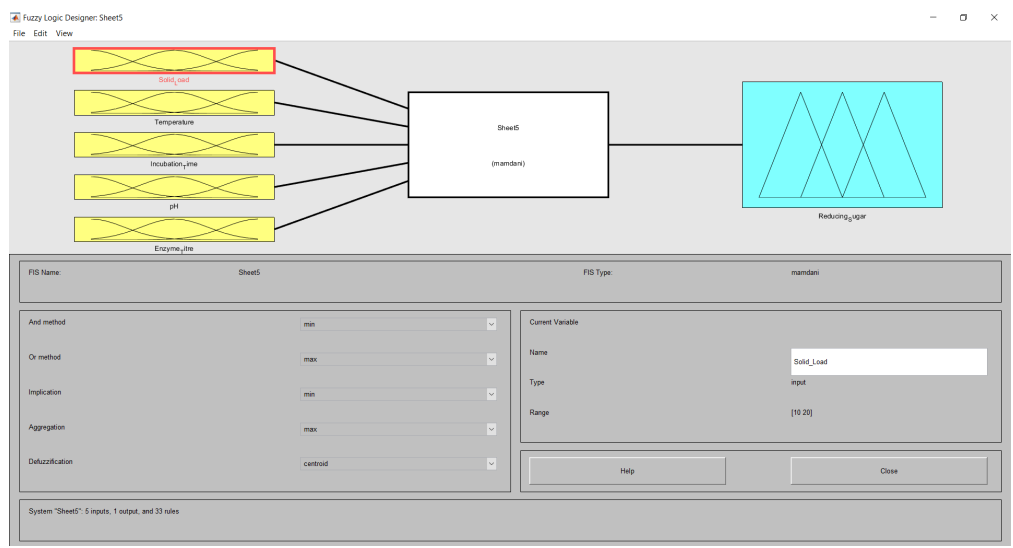

*Figure 7.1 Main Terminal With Inputs and Outputs*

The membership functions act as the domains or the sets where the values are stored and modelled by the system. The complexity of the algorithm can be adjusted based on the number of membership functions used. In simplistic terms a membership function can be thought of as an extension of the set theory which decides a range between which the 'fuzzy' values can be captured. The accuracy of the models can differ depending on the shape of the function that is chosen. However, as a general rule, gaussian functions are used due to their nature to better understand and adapt to process relationships. The number of functions chosen can reflect on the accuracy as well. However for inputs that only fall within a small range of values(such as pH), it may not always be useful to keep more functions if the values in experimentations aren't to extensive. If the maximal and minimal values are too spanned out, more membership functions should be added.

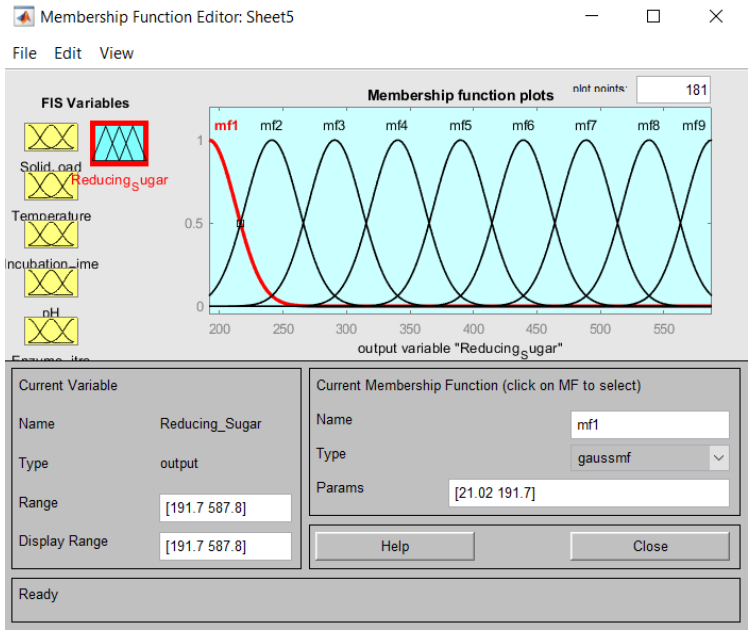

*Figure 7.2-Membership functions for the output with parameters and type.*

After creating the membership functions, rules are inserted into the algorithm based on a system of if-else statements. All statements with a specific input relate to a particular output. The rule creation for fuzzy logic models can be made levelled by introducing and/or conditions. The rules generated by the user will be related to the way the inputs relate to the output and can be modified to include outliers by changing the membership function ranges.

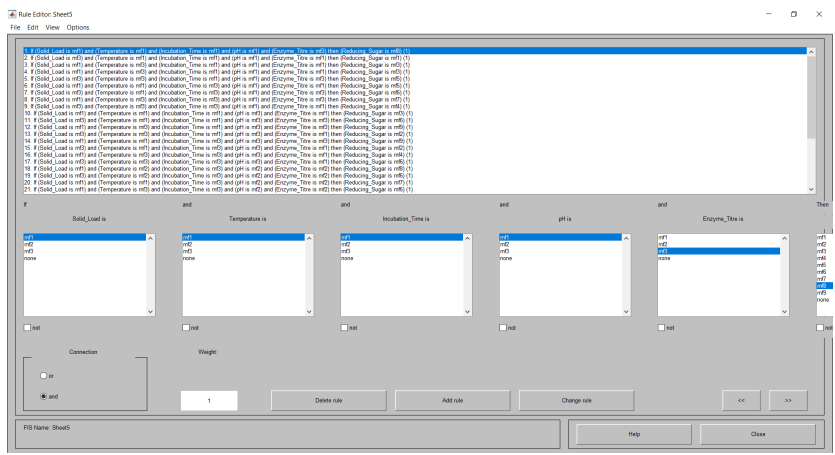

*Figure 7.3-Rules generated for the given dataset.*

On generating the rules, users can then view the model schema which shows how the system is learning the model and the output it is generating. In the case example discussed here, the reducing sugar value is obtained after the system interprets the rules fed through the interface. The yellow gaussian functions in the figure show the overlap between various inputs. To avoid redundancies, the system defuzzifies on a limit that is determined as per the number of rules used to generate the model. Creating more rules will thus lead to great accuracy and reduce repeating output values.

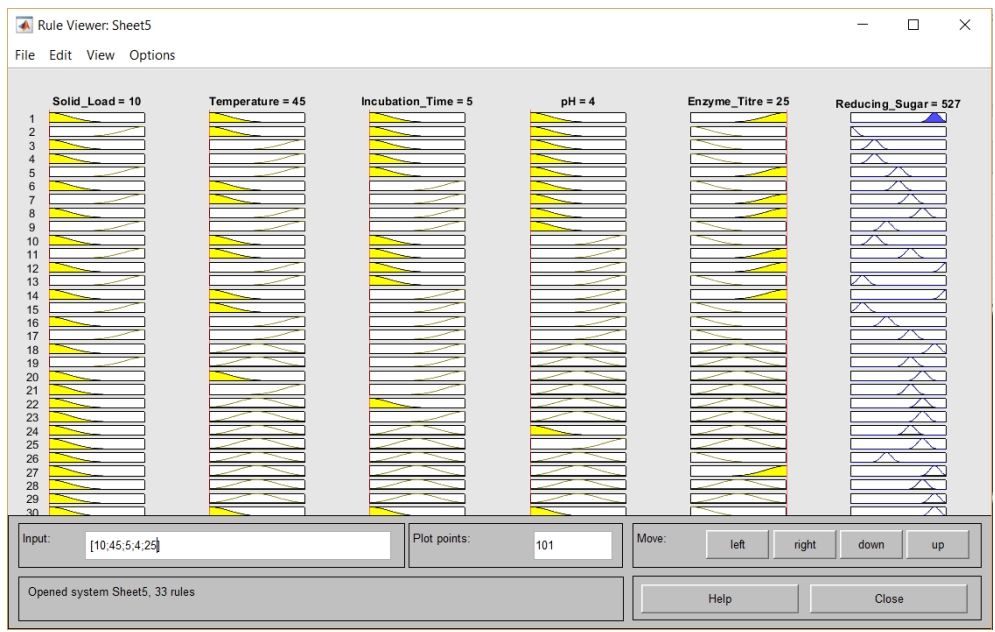

*Figure 7.4-Rules Generated with Input and Output Bars.*

A multidimensional plot for all the inputs and outputs can be generated to show the regions covered by the algorithm and how they vary with relation to each other. It is important to note that the lack of rules will only cause the system to generate output values based on previously defined rules resulting in errors. It is thus suitable to include a rule for every input so that the model can learn to incorporate and capture all sorts of values. A commonly known issue that arises in such model predictive control techniques is when an input value is entered that does not fit within the specified range articulated to the system. In the fuzzy logic modelling example, the user specifies the maximal and minimal range the input and output values can take which are still vulnerable to errors.

However, with the use of automated systems in industries, such models can be taught to 'update' these ranges when outliers arise so as to eliminate such inefficiencies. In the case that multiple outputs share inputs that fall within similar ranges, the system defines a cut-off so as to prevent redundant outputs from entering the model.

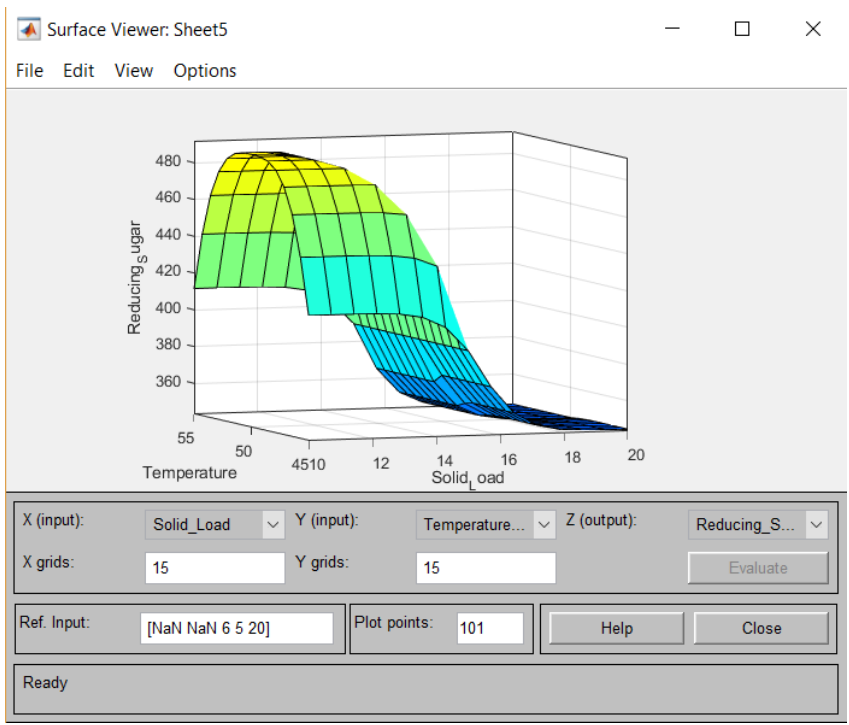

*Figure 7.5-Multidimensional plot for process parameters and output from MATLAB.*

## 8. MODEL RESULTS FOR CASES

For the fuzzy models generated, the model outputs have been plotted against the experimental outputs to illustrate how they can be used to model accurate predictive systems. The root mean square errors have been added to bring this to effect. The experiment number is represented on the y axis and the output is represented on the y-axis of the graphs below. The model outputs obtained from MATLAB have been compared against the actual outputs to show accuracies.

This also helps shed some light at how certain pitfalls can be avoided when modelling, in Fig.8.4, the high error can be offset by using better membership functions. While the error may be alarmingly high when compared to the small errors for the other case examples, it has been added to show how a lack of model rules can affect accuracies. On the contrary, higher accuracies may not always reflect incidents of overfitting as seen in Figure 9.2 and Figure 9.3. The model results for these figures show that the system has learnt to 'track' the values and can adapt well to new inputs. Within the scope of biofuel production, fuzzy enhanced sensors can become a reality in industries, moving away from pilot plants and small scale manufacturing facilities. While the crisp nature of the inputs and outputs may deter some from considering them as primary predictors in industries, they can still be used to test out relationship matrices in design experimentations to maximize yields and optimize operating conditions.

These results also derive consistency with the design methodologies discussed in literature that use dilute acids and various mixing mechanisms.( (Sohpal & Singh, 2015)

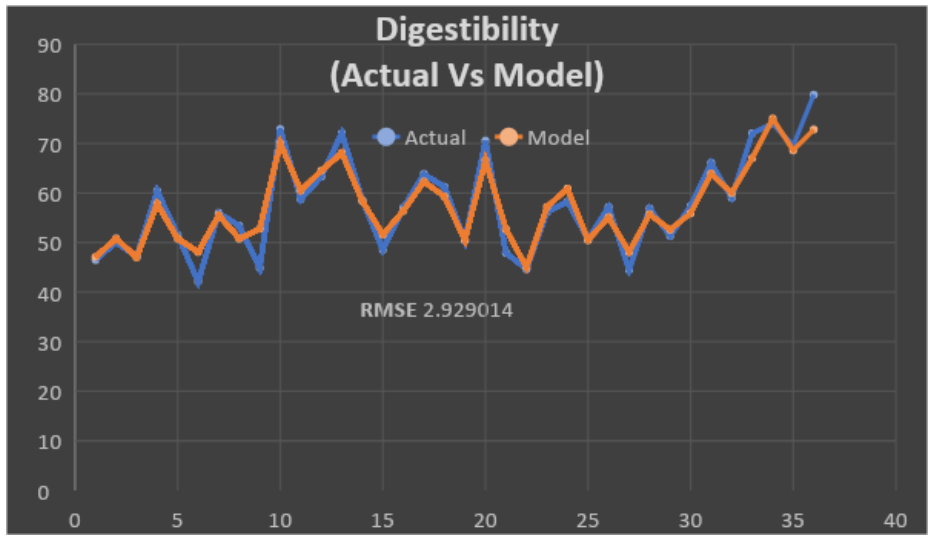

*Figure 8.1- Results from data obtained from (Rivera, et al., 2015)*

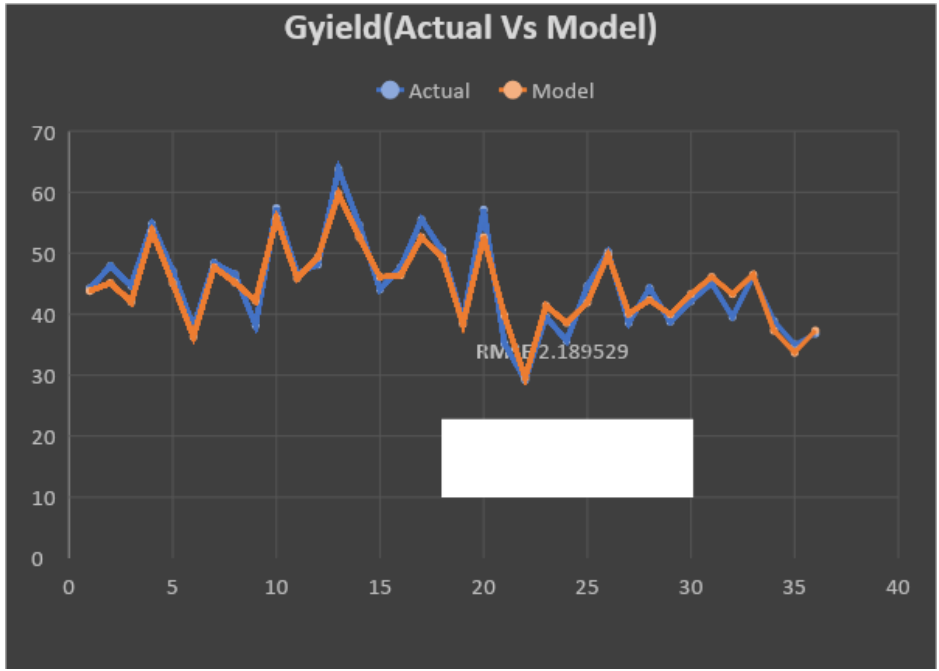

*Figure 8.2- Results from data obtained from (Rivera, et al., 2015)*

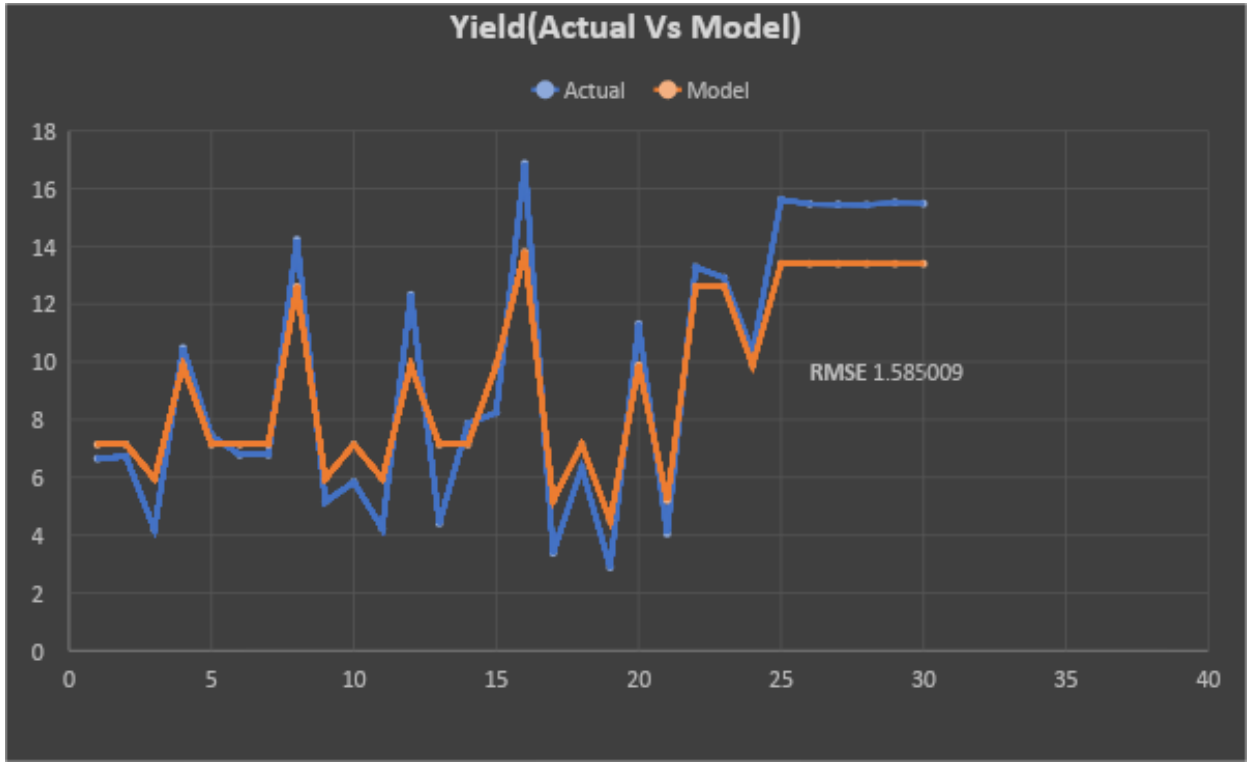

*Figure 8.3- Model accuracy for algorithm(Data obtained from Adnan, Suhaimi, Abd-Aziz, Hassan, & Phang, 2014)*

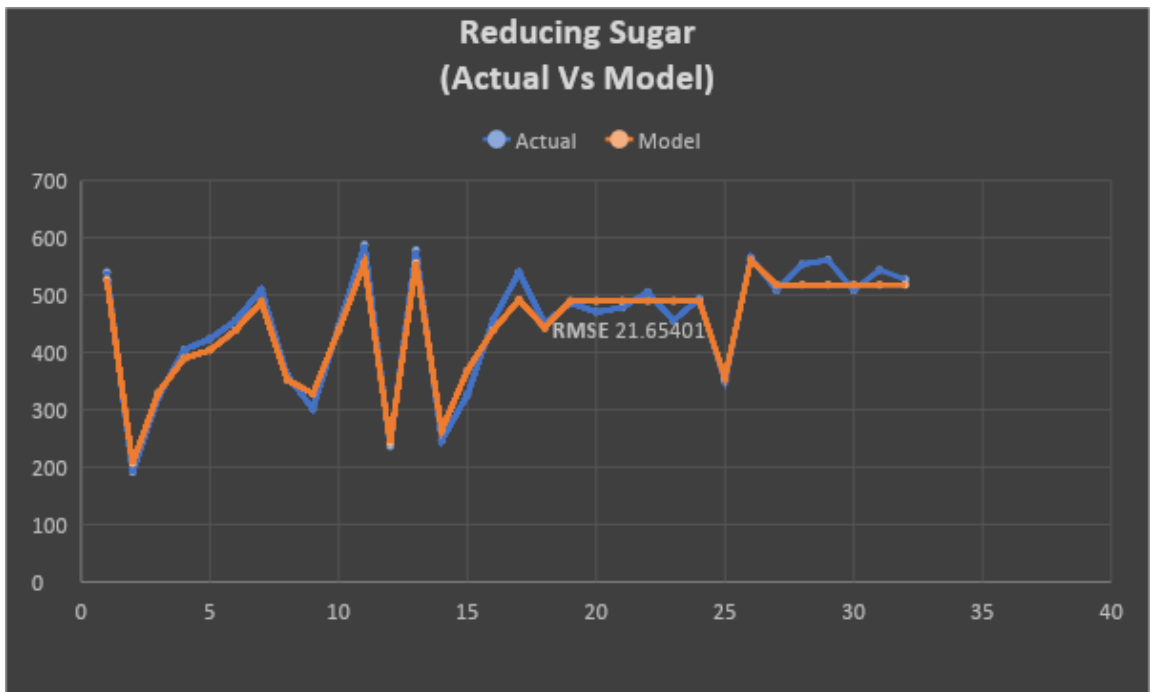

*Figure 8.4- Model accuracy for algorithm(Data obtained from Sherpa, Ghangrekar, & Banerjee, 2017)*

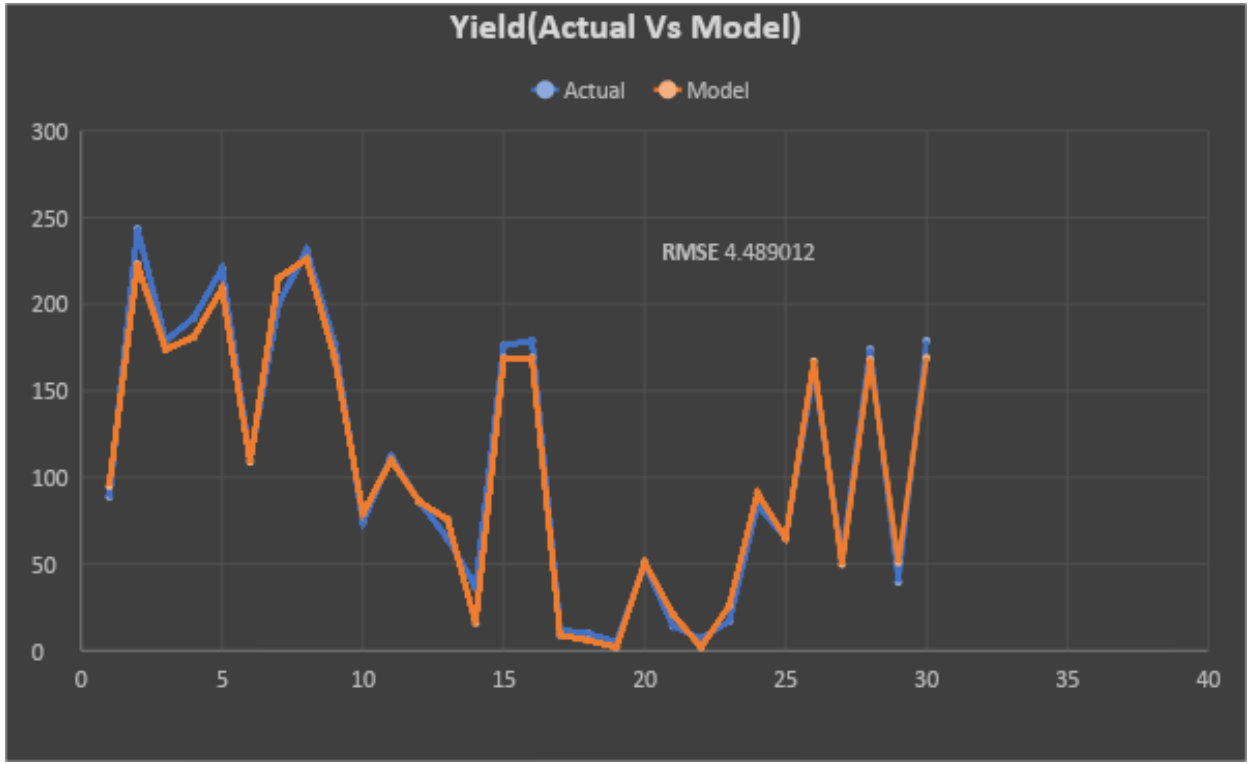

*Figure 8.5- Model accuracy for algorithm(Data obtained from El-Gendy, Madian, & Amr, 2013)*

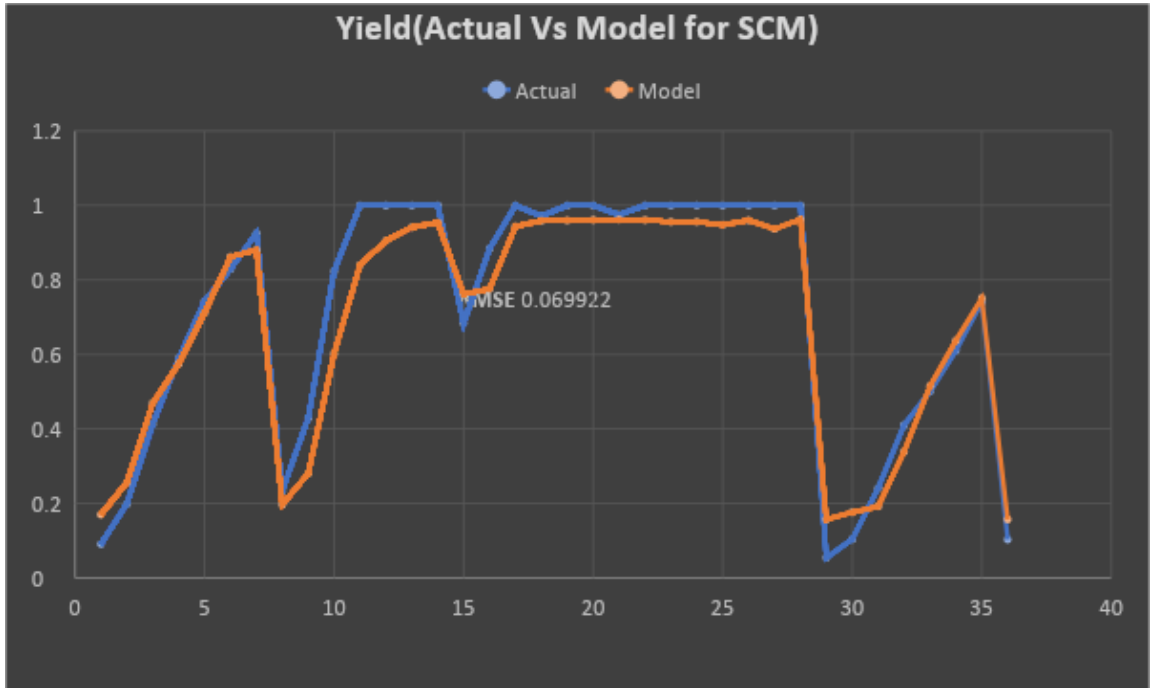

*Figure 8.6- Model accuracy for algorithm(Data obtained from Farobie, Hasanah, & Matsumura, 2015)*

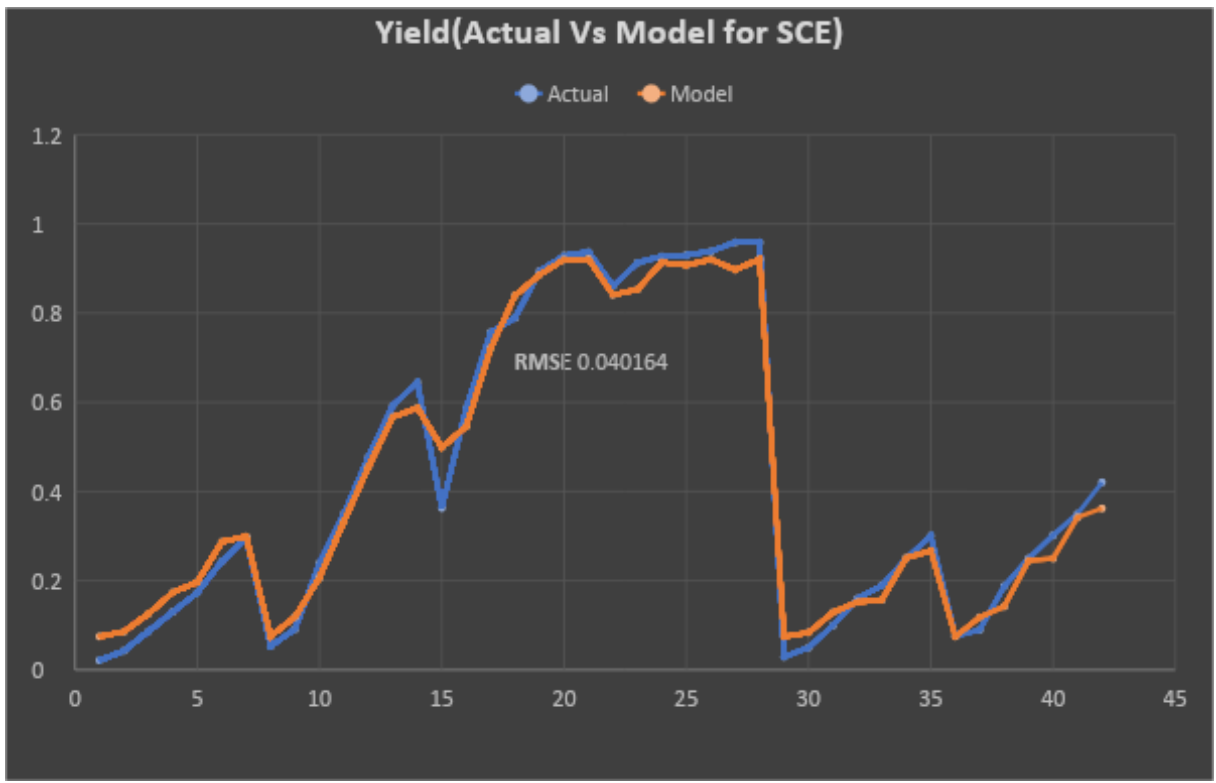

*Figure 8.8- Model accuracy for algorithm(Data obtained from Farobie, Hasanah, & Matsumura, 2015)*

The exceptional points in the plots where the expected output and model output have large differences are indications of the membership functions failing to properly capture the points to a greater accuracy which can be offset by creating more rules to make defuzzification easier.

**CONCLUSIONS**

The intricacies that affect the appropriate yields for ethanol and other biofuels tend to share a non-linear relationship with some of the inputs. Principal component studies of these parameters would indicate that they all can be modelled to fit equations using advanced levels of design studies such as FIS(Fuzzy Logic Interfaces). When used with a neuro-fuzzy design mechanism, the accuracy can be improved even further. The method of using design experimentations is quite crucial to industries as parameters that have significant influence on the yields can be identified. For fuzzy logic designs, this is quite important as the effects of variables must be included in a quantifiable way. The results obtained from parametric designs from (Timung, Naik Deshavath, Goud, & Dasu, 2016) and (Mohan, Ramesh, & Reddy, 2012) show that removing certain variables can have a better influence on the models that can predict the values with greater accuracy.

As per the fuzzy logic models that have been used here, the conclusion to be made is that specifying the nature of the functions for the inputs as well as the ranges can produce accurate models to great degrees. Preliminary analysis too can help understand how the variables are related so as to frame the rules in more dynamic ways. Certain models show great accuracies such as the one developed for the data in Adnan, Suhaimi, Abd-Aziz, Hassan, & Phang, 2014.

While others may show great error rates, these can be reduced by creating better itnerface rules and making more membership fucntions that can correlate the inputs and outputs better. CCD and RSM methodologies are particularly of great interest to the industry as they help develop experimentations from a large batch of runs and focus on the ones that promise great yields. In the case examples discussed in this paper. A majority of the experimentations take use of modelling studies with a predictive approach which shows how tools like fuzzy modelling can help industries meet product objectives by optimizing inputs. Biofuel demands will likely see great increases in the future owing to the transient nature of conventional fuels and the risks they pose. As a result, model predictive analogies such as fuzzy logic will become popular to better understand the process and help industries meet optimal production levels.